\documentclass[twocolumn,showpacs,preprintnumbers,amsmath,amssymb]{revtex4}
\usepackage{graphicx}
\usepackage{dcolumn}
\usepackage{bm}
\newcommand{\be}{\begin{eqnarray}}
\newcommand{\ee}{\end{eqnarray}}

\begin{document}
\title{Wave Propagation in the Cantor-Set Media: \\
  Chaos from Fractal}
\author{Kenta Esaki, Masatoshi Sato, and Mahito Kohmoto}
\affiliation{%
Institute for Solid State Physics,
Kashiwanoha 5-1-5, Kashiwa, Chiba, 277-8581, Japan
}%
\date{\today}
\begin{abstract}
Propagation of waves in the Cantor-set media is investigated by a
renormalization-group type method. We find fixed points for complete
reflection, $T^*=0$, and for complete transmission, $T^*=1$. 
In addition, the wave numbers for which
transmission coefficients show chaotic behaviors
are reported.  
The results obtained are for optical waves, 
and they can be tested in optical experiments.
Our method could be applied to 
any wave propagation through the Cantor set.
\end{abstract}


\maketitle


Localization of the electronic states due to disorder is an active field in
condensed matter physics. It has been recognized that localization
occurs not only in disordered systems but also in quasiperiodic
systems in one dimension \cite{MK83,KO84,MK86}.
While the localization of states was originally regarded as an
electronic problem, it was later recognized that the phenomenon is
essentially a consequence of the wave nature of electronic
states. Therefore, such localization can be expected for any wave
phenomenon. An optical experiment with the Fibonacci multilayer was
proposed\cite{mk87}, and the corresponding experiments using dielectric
multilayer stacks of SiO$_2$ and TiO$_2$ thin films were
reported\cite{wg94,th94}. The transmission coefficients and
scaling properties well agree with the theory\cite{mk87}, and can be
considered as an experimental evidence for localization of
electromagnetic waves.
 Recently, an experiment for propagation of electromagnetic waves in
a three-dimensional fractal cavity was reported\cite{mw04}, where a
confinement of electromagnetic waves in the fractal structure was
observed.

Propagation of light in the Cantor media was numerically studied by
Konotop et al.\cite{ko90} and Sun and Jaggard \cite{xs91},
 later by Bertolotti et al. with transfer matrices\cite{mp94,mb96},
 and recently by Yamanaka and Kohmoto\cite{nh05}
 and Hatano\cite{nh06} again with transfer matrices.
Comparison of the experimental results and numerical results was performed
by Lavrinenko et al.\cite{av02}.

We study wave propagation in the Cantor-set media by 
a renomalization-group type method\cite{kadanoff}.
For specific values of the wave numbers,
 transmission coefficients show chaotic
behaviors governed by the logistic map.
This exotic behavior could be observed in an optical experiment.
In this system, the one-dimentional
theory is strictly valid. In addition, it is feasible to construct
the system accurately, and the parameters can be precisely 
controlled and measured.

\begin{figure}
 \begin{center}
  \includegraphics[width=4.5cm,clip]{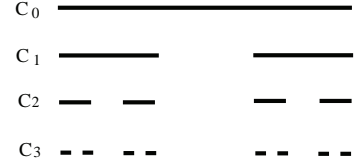}
\caption{\label{fig:cantorsequence}The initial state $j=0$ and the first
  three generations in the construction of the Cantor set.}
\end{center}
\end{figure} 
The procedure of constructing the Cantor set begins with a line segment
with unit length. We regard this as substrate A. The line segment is
divided into three parts. The left and the right segments are
substrate A, each of which has length ${1}/{3}$. 
The middle part, which has length ${1}/{3}$, is removed.
We regard the removed part as substrate B.
The procedure is repeated for each of line segments A. 
We call $j$-th generation of the Cantor sequence as C$_j$.
For C$_j$ we have a set
of $2^j$ line segments of substrate A, each of which has length ${1}/{3^j}$.
The Cantor set is obtained if this procedure is repeated infinite times.
It is self-similar and has the fractal dimension
$\log{2} /\log{3}$. 
Constructions of first few generations are shown in
Fig.\ref{fig:cantorsequence}.
In the following, we suppose that the indices of refraction 
of A and B are $n_{\rm A}$ and $n_{\rm B}$,
and the magnetic permeabilities of A and B
are $\mu_{\rm A}$ and $\mu_{\rm B}$, respectively.

\begin{figure}
 \begin{center}
  \includegraphics[width=5.5cm,clip]{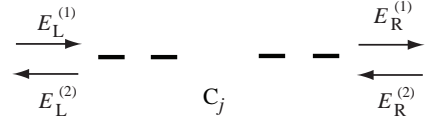}
\caption{\label{fig:interface}
  Electromagnetic wave propagation through the Cantor sequence C$_j$.} 
\end{center}
\end{figure}

We consider wave propagation through C$_j$. See Fig.\ref{fig:interface}.
For simplicity, suppose that the incident light is normal
and the polarization is perpendicular to the plane of the light path.
Here $E_{\rm L}^{(1)}$ represents the incident light, 
and $E_{\rm L}^{(2)}$ and $E_{\rm R}^{(1)}$ represent 
reflected and transmitting light, respectively.
We also consider light coming from right $E_{\rm R}^{(2)}$.
By setting
\begin{equation} 
E_+=E^{(1)}+E^{(2)},   
\quad
E_-=(E^{(1)}-E^{(2)})/i,
\end{equation} 
 the wave propagation is given by\cite{mk87,nh05}
\begin{eqnarray}
\left(
\begin{array} {c} 
 E_+ \\ 
 E_-
\end{array}
\right)_{\rm R}
  ={\cal M}_j (k)
  \left(
\begin{array} {c} 
E_+\\
E_- 
\end{array}
 \right)_{\rm L},
\label{ET}
\end{eqnarray}
 where ${\cal M}_j$ satisfies
\begin{equation}
 {\cal M}_{j+1}(k)={\cal M}_j(\frac{k}{3} ) 
{\cal T}(\frac{n_{\rm B}k}{3}) 
{\cal M}_j(\frac{k}{3}), 
\label{MT}
\end{equation}
with the initial condition 
\begin{eqnarray}
  {\cal M}_0(k)={\cal T}_{\rm BA} {\cal T}(n_{\rm A}k) 
{\cal T}_{\rm AB}.
 \label{M0}    
\end{eqnarray}
Here $k$ is a wave number in vacuum,
and the matrices ${\cal T}_{\rm AB}$ and ${\cal T}_{\rm BA}$ represent light
propagation across interfaces $A \gets B$ and $B \gets A$
respectively,
\begin{eqnarray}     
{\cal T}_{\rm AB}= 
{\cal T}_{\rm BA}^{-1}=
\left(
\begin{array} {cc}
1 & 0  \\
 0 & 1/n
\end{array} 
\right),
\quad
n = \frac{n_{\rm A}}{n_{\rm B}} \times \frac{\mu_{\rm B}}{\mu_{\rm A}}.
\end{eqnarray}
The matrix ${\cal T}(\delta)$ represents light propagation
within a layer,
\begin{eqnarray}
{\cal T}(\delta)=
\left(
\begin{array} {cc} 
 \cos\delta & -\sin\delta \\ 
  \sin\delta & \cos\delta 
\end{array}
 \right).
\label{transfer}
\end{eqnarray}
For a layer of type A and B with thicknesses $d$, 
the phase $\delta$ is given by
$\delta=n_{\rm A} k d$ and $\delta=n_{\rm B} k d$,
respectively. 
Equation (\ref{MT}) can be regarded as a
renormalization-group transformation.
One can easily show that
\be
 \det {\cal M}_j = 1.
\label{detm}
\ee

By putting $E_{\rm R}^{(2)}=0$, the transmission coefficient $T_j$ of
 ${\rm C}_j$ is given by
\begin{equation}
T_j=\frac{4}{|{\cal M}_j|^2+2}, 
\label{MJ}
\end{equation}
where $|{\cal M}_j|^2$ is the sum of the squares of each matrix
element of ${\cal M}_j$.
\begin{figure}
 \begin{center}
  \includegraphics[width=4.6cm,clip]{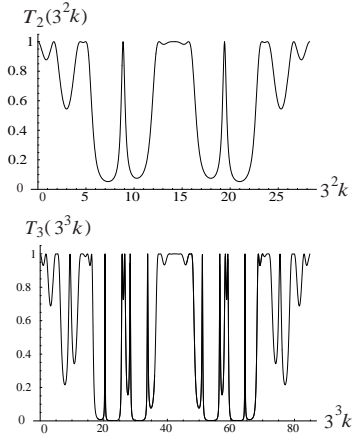}
  \caption{\label{fig:transmission}Transmission coefficient $T_j$ as a
  function of the wave number $k$ for C$_2$ (top) and C$_3$ (bottom).
 ($n_{\rm A}=2$, $n_{\rm B}=1$ and $n=2$.) The range of the wave number $k$ for C$_j$ is $0 \le k \le 3^j \pi$. Note that the
  horizontal axes are rescaled.} 
 \end{center}
\end{figure} 
An example of transmission coefficients as a function of
 wave numbers is shown in Fig.\ref{fig:transmission} for C$_2$ and C$_3$. 
They show a scaling behavior when one multiplies wave numbers by three 
 each time one increases a generation.
Thus we study wave propagation of wave numbers $3^j k$ of C$_j$.

In C$_j$, all A's have the same length $1/3^j$,
 thus the propagation in A's are given by the unique transfer matrix
${\cal T}(n_{\rm A} k)$.
On the other hand, B's have lengths 
$1/3^j, 1/3^{j-1}, \cdots$ and $1/3$,
thus in general they give $j$ different transfer matrices,
 which are given by ${\cal T}(n_{\rm B} k), {\cal T}(3 n_{\rm B} k), \cdots$,
 and ${\cal T}(3^{j-1} n_{\rm B} k)$.
However, if $k$ is given by
\begin{eqnarray}
n_{\rm B}k=\frac{m \pi}{3^{q}},
\label{coherentk}
\end{eqnarray} 
with integers $m$ and $q$,
 ${\cal T}(\delta)$'s in large B's 
are equal to ${\cal T}(m \pi)$.
Now the map (\ref{MT}) for $j\ge q$ 
is written as
\begin{equation}
{\cal N}_{j+1}={\cal N}_j^2,
\label{NT}
\end{equation}
where
\be
{\cal N}_j =
{\cal T}(m \pi){\cal M}_j.
\ee
From (\ref{detm}), ${\cal N}_j$ satisfies
\begin{eqnarray}
 {\rm det} {\cal N}_j=1. 
\label{detn}
\end{eqnarray}
To solve (\ref{NT}), we rewrite ${\cal N}_j$ in terms of the Pauli matrices $\sigma_i$ $(i=1.2.3)$,
\begin{eqnarray}
{\cal N}_j
=&x_j{\bm 1}+{\bm a}_j\cdot {\bm \sigma}, \quad {\bm a}_j=(a_j^{(1)},ia_j^{(2)},a_j^{(3)}),
\end{eqnarray}
where $x_j$ and $a_j^{(i)}$ $(i=1,2,3)$ are real.
Comparing the coefficients of ${\bm 1}$ and ${\bm \sigma}$ in both side of (\ref{NT}), we have
\be
x_{j+1}=x_j^2+{\bm a}_j^2,
\label{xja}
\ee
and
\be
{\bm a}_{j+1}=2 x_j {\bm a}_{j}.
\label{aaj}
\ee
In addition, (\ref{detn}) leads to
\be
x_j^2-{\bm a}_j^2=1.
\label{detn2}
\ee
From (\ref{xja}) and (\ref{detn2}), we have the map for
 $x_j=\frac{1}{2}{\rm Tr} {\cal N}_j$
\begin{equation}
x_{j+1}=2x_j^2-1,
\quad
(j\ge q).
\label{xjlogistic}
\end{equation}
This is the logistic map which maps the interval $[-1,1]$ onto itself
and others to $+\infty$. The bounded $x_j$'s are given by
\begin{equation}
x_j=\cos[2^{j-q}(\cos^{-1}{x_q})],
\end{equation}
and have chaotic behaviors with the Lyapunov exponent $\ln 2$.
\begin{figure}
 \begin{center}
   \includegraphics[width=4.5cm,clip]{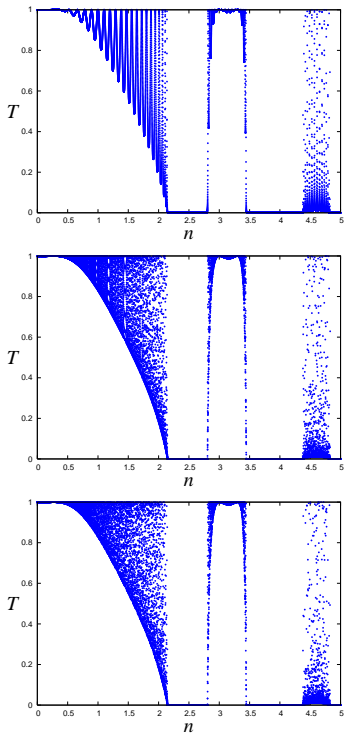}
   \caption{\label{fig:chaos}Transmission coefficient $T_j$ as a
   function of $n$ for $n_{\rm B} k=\pi/3$.
 Iteration number is $j=5$ (top), $j=10$ (center), and $j=100$ (bottom).} 
 \end{center}
\end{figure}
From (\ref{aaj}), we also find the following
constants of motion $\lambda$ and $\eta$
\begin{eqnarray}
\frac{a_j^{(3)}}{a_j^{(2)}}=\lambda,
\quad
\frac{a_j^{(1)}}{a_j^{(2)}}=\eta
\quad (j\ge q).
\label{const12}
\end{eqnarray}
In terms of them, the transmission coefficient (\ref{MJ}) is given by 
\begin{eqnarray}
T_j=\frac{1-(\lambda^2+\eta^2)}{1-(\lambda^2+\eta^2)x_j^2}.
\label{logistic_T} 
\end{eqnarray}
If the constants of motion $\lambda$ and $\eta$ satisfy
 $0 \le \lambda^2+\eta^2\le 1$, we have $|x_j| \le 1$.
Then, the complete transmission $T_j=1$ is achieved when $x_j=\pm 1$,
and we have the minimal transmission
\begin{eqnarray}
T_{\rm min}=1-(\lambda^2+\eta^2),
\end{eqnarray}
when $x_j=0$.
On the other hand, if ${\lambda}^2+{\eta}^2 >1$ we have $|x_j|> 1$.
Then $x_j$ escapes to infinity
and the transmission coefficient flows into the fixed point for
complete reflection $T^*=0$, as seen from (\ref{logistic_T}). 
We show the transmission coefficient $T_j$ as a function of $n$
for $n_{\rm B} k=\pi/3$ in Fig.\ref{fig:chaos}. 
Since the complete reflections and the 
complete transmissions are nearby in $n$, 
 the Cantor media could be used as a fast swiching devices.

We note that similar chaotic behaviors also appear for the initial wave numbers
\begin{eqnarray}
n_{\rm B}k=\frac{m+1/2}{3^{q}}\pi,
\label{coherent2}
\end{eqnarray}
with integers $m$ and $q$.
The matrix ${\cal T}(\delta)$ in large B's
is now ${\cal T}((m+1/2)\pi)$, and
the map for ${\cal N}_j$ is
${\cal N}_{j+1}=-{\cal N}_j^2$.
The trace of ${\cal N}_j$ follows the logistic map, again.
\begin{figure}
 \begin{center}
   \includegraphics[width=4.5cm,clip]{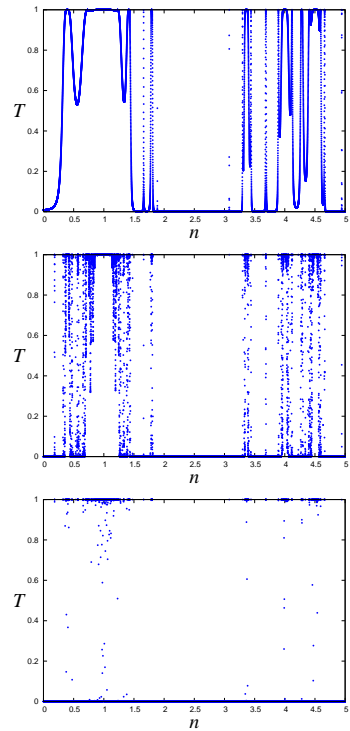}
   \caption{\label{fig:nonchaos}Transmission coefficient $T_j$ as a
   function of $n$ for $n_{\rm B} k=\pi/4$.
 Iteration number is $j=5$ (top), $j=10$ (center), and $j=100$ (bottom).} 
 \end{center}
\end{figure}
\begin{figure}
 \begin{center}
  \includegraphics[width=4.5cm,clip]{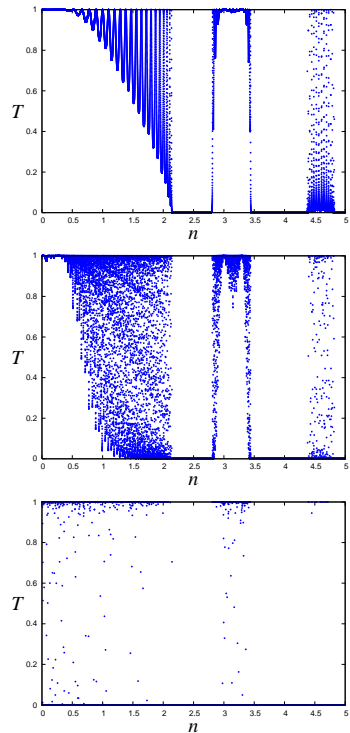}
  \caption{\label{fig:nearchaos}Transmission coefficient $T_j$ 
as a function of $n$ for $n_{\rm B} k=\pi/3+0.0001$.
Iteration number is $j=5$ (top), $j=10$ (center), and $j=100$ (bottom).} 
 \end{center}
\end{figure}

 For initial wave numbers other than
(\ref{coherentk}) or (\ref{coherent2}),
 our numerical calculations suggest that light reflects completely
 or transmits completely by the Cantor-set media.
An example of our numerical results are shown 
in Fig.\ref{fig:nonchaos}.
The transmission coefficients tend to 
flow into the fixed point $T^*=0$ or $T^*=1$
 as the generation increases.

When initial wave numbers are near a value 
(\ref{coherentk}) or (\ref{coherent2}),
 the transmission coefficients show chaotic behaviors 
in first few generations.
For example, the transmission coefficients $T_j$ for $n_{\rm B} k=\pi/3+0.0001$
are shown in Fig.\ref{fig:nearchaos}.
In the fifth generation ($j=5$), the transmission coefficients show
 chaotic behaviors indistinguishable from the case for 
$n_{\rm B} k=\pi/3$ (Fig.\ref{fig:chaos}).
However, in the tenth generation ($j=10$),
 the transmission coefficients show different behaviors 
 from the case for $n_{\rm B} k=\pi/3$,
 and in the hundredth generation ($j=100$),
 they tend to flow into the fixed point $T^*=0$ or $T^*=1$.
 Our numerical results again suggest that light reflects completely or
 transmits completely by the Cantor-set media.
This behavior should be observed experimentally.

We can also study wave propagation 
for other types of Cantor-set media by using 
similar recursion relations as (\ref{MT}).    
 Chaotic behaviors governed 
by the logistic map occur in the Cantor-set media,
if the length of each A becomes $1/\alpha$ each time
 one increases a generation, and $\alpha$ is an integer.
Details will be given in a separated paper \cite{ESK}.

We acknowledge useful discussions with M. Yamanaka.

\end{document}